# Assessing m-Learning Adoption in Higher Education

Muasaad Alrasheedi
ITC Department
Arab Open University
Jeddah, Saudi Arabia
m.alrasheedi@arabou.edu.sa

Luiz Fernando Capretz
Department of Electrical & Computer Engineering
Western University
London, Canada – N6A5B9
lcapretz@uwo.ca

*Abstract*—New mobile platforms, connected seamlessly to the Internet via wireless access have become increasingly more powerful and have found usage in a diverse set of application areas, including the education sector. The educational institutions are becoming more open to embracing new learning platforms, which in turn has sparked the interest in developing new assessment frameworks. This paper has used the framework of the Capability Maturity Model (CMM) to design a model for M-learning within educational institutions. The framework has been validated with studies cases from higher education institutions.

I. INTRODUCTION

Mobile devices have become ubiquitous in the present day world because of their versatility and hence capability to be used in several different ways. The use of devices like mobile phones and more recently tablet computers, have found their way into the educational sector as well. Learners and educators around the world are increasingly using the mobile technology to access information, streamline administration and facilitate learning in new and innovative ways. Many years ago senior executives in a company brought personally owned mobile devices to work and requested the office IT department to support it, giving birth to the Bring Your Own Device (BYOD) trend [1].

Mobile leaning or m-Learning is a learning platform that provides learners 'anytime-anywhere access to educational and university resources. Learners find themselves empowered by using mobile technology to gain access to the required course materials even when they are disconnected from the network. Because of the immense advantages of mobile learning, not to mention the enormous untapped market comprising of most of the universities of the world, the area of m-Learning is also seen as an extremely lucrative business opportunity.

The first objective of this paper is to address the benefits that educational institutions will get at various stages of their adopting the m-Learning platform for imparting education. The second goal of the research is to present a maturity model [2], [3], [4] to assess the adoption of m-learning in institutions of higher education based on the well-known Capability Maturity Model (CMM). An M-Learning Maturity Models (MLMM) was developed, evaluated, and introduced in the paper.

II MOBILE LEARNING MATURITY MODEL (MLMM)

The five levels of the MLMM are listed below. For each level, different statements based on CSFs must be satisfied to achieve maturity at that level [5].

*A. First Level: Preliminary*

At this level, universities and the institutions do not consider mobile devices to be important in the provision of their services and products. It is a reactive and experimental stage that recognizes the need to improve provision of information through mobile devices. The institute lacks vision and measures for implementation of the prototype. A number of limitations exist, like the limited coverage of mobile devices and limited understanding of m-Learning's value by students.

*B. Second Level: Established*

This level is based on the recognition of the opportunity provided by mobile devices in the education system and, particularly, in blended learning. This stage is characterized by interest from students and instructors resulting in investment in m-Learning technologies by the institute. Thus, while institutes guide and facilitate m-Learning implementation, they lack the ability to evaluate the m-Learning systems. This brings the need for improvements in the existing and implemented pilot prototypes using the nine CSFs to achieve m-Learning maturity to this level.

*C. Third Level: Defined*

In the defined level, mobile device is considered critical for the interaction among students, instructors, and administrative staff. In order to succeed at this level, the learning institutions must link their m-Learning strategies with core and technical visions; invest heavily in the system; and have a well-defined change management plan to carry out the m-Learning transition.

*D. Fourth Level: Structured*

In the structured level, m-Learning is characterized by optimization and innovation. The optimization results in a rich, dynamic, and flawless experience for students and instructors in using m-Learning system. The University uses techniques to refine procedures and policies to control any changes experienced in m-Learning that help and increase students and instructors engagement. The use of mobile device applications allows students to provide feedback, give comments, and share information. As a result, institutions refine and improve procedures and policies to control any changes experienced in mobile technology.

*E. Fifth Level: Continuous Improvement*

Finally, the highest m-Learning maturity level is the continuous improvement level. In this stage, m-Learning has already been accepted as the best approach to provide knowledge and exchange of information between students and instructors. In this stage, institutions are constantly evaluating themselves to ensure continuous improvement and optimization. This helps to identify any changes that might limit or change the manner in which m-Learning is used [6].

III RATING METHODOLOGY

We have used terms such as m-Learning factors Rating (mLRt), Number of Achieved Statements (NAS), Passing Threshold (PT), and m-Learning Maturity Level (MLML). In the statistical equation for our maturity model, the following abbreviations and symbols are used:

MF = m-Learning Factor

MFN = M-Learning Factor Number (an integer)

ML = Maturity Level (an integer)

S = Statement

SN = Statement Number (an integer)

NAS = Number of Achieved Statement

Let $MF_t[i, j]$ be a rating of the ith CSFs of the jth maturity level. Subsequently, according to the scales, it can be summarized as:

$mLR_t[i,j] = 4$, if the Achievement of the condition / statement is at least 80%

$= 3$, if the Achievement of the condition / statement is from 66.7 to 79.9%

$= 2$, if the Achievement of the condition / statement is from 33.3 to 66.6%

$= 1$, if the Achievement of the condition / statement is less than 33.3%

$= 0$, if the condition - statement is not applicable.

An ith condition/statement at the jth maturity level is considered Achieved if $mLRt[i, j] \geq 3$ or $mLRt[i, j]$ is 0. The number of conditions/statements Achieved at jth maturity level is defined as:

$$NAS[j] = Number\ of\ \{mLR_t[i,j] \geq |Achieved\}$$

$$= Number\ of\ \{mLR_t[i,j]\ |\ mLR_t[i,j] \geq 3\ or\ mLR_t[i,j] = 0\}$$

The m-Learning maturity is considered to be achieved if 80% of the conditions or statements in the questionnaire are achieved. Thus, if TNS [j] is the Total Number of Statements at the jth maturity level, then the passing threshold (PT) at the jth maturity level is defined as:

$$PT[j] = TNS[j] * 80\ \%$$

IV CASE STUDIES

According to Flyvbjerg [7], "The case study is useful for both generating and testing of hypotheses but is not limited to these research activities alone." Thus case study can be helpful in the beginning steps of a study by providing hypotheses which can be experimented scientifically.

*A. Evaluation Process*

Our model was applied to two m-Learning programs in two universities (*country name removed*) to perform the m-Learning maturity assessment. The universities will be referred to as "University A" and "University B," to protect their privacy. Using a Likert scale ranging from 0 to 4, the participants were requested to provide the degree of agreement with each statement for the questionnaires designed [(0 Inapplicable), (1 Unachieved), (2 Partially Achieved), (3 Largely Achieved) and (4 Completely Achieved)]. Consequently, the questionnaire was completed by the participants starting from Level 2 and finishing at Level 5.

The respondents of this study included the Dean, higher management staff, or a faculty member. Survey link (SoGoSurvey tools) and email were the means of all communication with the respondents. The participants in the study had consented to their involvement and they were not paid any reimbursement. In the following sections, both case studies are discussed. Bias in the sample is limited because multiple responses were received from each university. A more accurate description of the m-Learning was provided by different respondents. Inter-rater agreement analysis has also been performed and the degree of agreement among all the raters within each university is known and provided information. The following section describes the analysis.

*B. Case Study – "University A"*

University "A" has a Blackboard system and we received a total of 8 complete responses from the university. As proposed by the rating method conferred in Section above (IV-B), if the performance scale is larger than or equal to 3 or 0, then statements are considered to be agreed upon. We have calculated NAS (the Number of Achieved Statements) for all the levels. NAS is 14 for Level 2, 15 for Level 3, 1 for Level 4, and 1 for Level 5 from the data collected. NAS at Level 2 has a pass limit of 80% according to the rating limit for our MLMM. University (A) is therefore at the "Established" maturity level. As the value of the statement is 3, it is considered that level 2 is largely achieved.

*C. Case Study – "University B"*

University "B" has its own Learning Management System and we received a total number of 8 complete responses from the university. According to the rating method discussed in Section IV above, if the performance scale is larger than or equal to 3 or 0, then a statement is considered to be agreed upon. We have calculated NAS (the Number of achieved statements) for all the levels. NAS is 17 for Level 2, 2 for Level 3, 0 for Level 4, and 0 for Level 5 from the data collected. NAS at Level 2 has a pass limit of 80% according to the rating limit for our MLMM. University (B) is therefore at the "Established" maturity level.

Summarized assessment results for both case studies are given in Table I.

TABLE I.    SUMMARY OF ASSESSMENT RESULTS OF CASE STUDIES

| MLML | Total Questions | Pass Threshold 80% | Un. "A" NAS | Un. "B" NAS |
|---|---|---|---|---|
| Preliminary | 0 | Not Valid | - | - |
| Established | 18 | 14 | 14 | 17 |
| Defined | 20 | 16 | 15 | 2 |
| Structured | 20 | 16 | 1 | 0 |
| Continuous Improvem. | 17 | 14 | 1 | 0 |

*D. Analysis of Inter-Rater Agreement*

The extent of agreement between different raters within one university is provided by inter-rater agreement. The assessment of the identical methodologies adheres to inter-rater agreement and conforms to reproducibility. In cases where data is ordinal, the Cohen's Kappa [8] is preferred to evaluate inter-rater agreement.

An inter-rater agreement analysis has been conducted in our study using Kappa statistics. 17 respondents participated – 8 from university A and 9 from university B. The values of Cohen's Kappa and the Fleiss Kappa coefficients can range from 0 to 1, with 0 indicating perfect disagreement and 1 indicating perfect agreement. In University A, the benchmark for Kappa does include four level scales, where $< 0.44$ is poor agreement, $0.44$– $0.62$ is moderate agreement, $0.62$–$0.78$ is substantial agreement, and $> 0.78$ is excellent agreement. For University A, the Kappa coefficients range from 0.45 to 0.85 and therefore, are classified as moderate agreement.

Likewise, the Kappa coefficients ranged from 0.43 to 0.79 when we did same analysis for university B. Thus, in the case of university B, coefficients are classified as being moderate agreement too.

V DISCUSSION

In software engineering, maturity model information about different processes is provided including their current maturity levels and their related activities. An organization can seek help from this information to upgrade their processes, plan their future activities, and design strategic plans. End user experiences can provide great help to improve the software projects. Consequently, m-Learning and correlated problems are the key areas of study in academic society. Assessment is needed to determine particular areas where improvements are compulsory.

M-Learning is a relatively new disciplinary research area, and m-Learning adoption requires a comprehensive strategy due to its continuous adoption. In previous work that we have done, we examined different key factors of m-Learning adoption. The significant key factors are the measuring instrument to introduce an MLMM in the assessment methodology for m-Learning. The structural MLMM composition consists of the evaluation framework from three dimensions relying on university management approach, and on students and instructors.

Consequently, the current maturity of an m-Learning platform is assessed by this model with assessment methodology of defining and conducting case studies. An integral feature of the MLMM is the methodology for specifically evaluating m-Learning platform maturity. This model will help university management perform adoption assessments for their m-Learning projects and boost their upgrading strategies.

VI CONCLUSION

Our MLMM is based on nine key factors, and we have empirically analyzed and identified them in the three previous studies. The area that is less attractive to the researchers is the CSF assessment of m-Learning, and, accordingly, a process that estimates the m-Learning maturity is the main contribution of this work. An evaluation questionnaire for four of the five maturity levels is part of composition of the framework of this model, as well as a rating methodology and a performance scale. Additionally, we have also studied the execution of two m-Learning projects in two universities and discussed the findings as case studies. Leaving the limitations aside, this work has contributed to setting up an all-inclusive approach for m-Learning maturity and addressed the imperative subject of factors of evaluation in m-Learning.